\documentclass[12pt]{article}
\usepackage{graphicx}
\usepackage{amsmath,amssymb,amsfonts}
\usepackage[numbers,sort&compress]{natbib}
\usepackage[colorlinks=true, linkcolor=blue, urlcolor=blue, citecolor=blue]{hyperref}
\usepackage{bm}
\usepackage{tabularx}
\usepackage{xcolor}
\usepackage{booktabs}
\usepackage{caption,subcaption}
\usepackage{authblk}
\usepackage{tabularx}

\title{\textbf{Cosmology of $f(Q,L_m)$ gravity with Holographic Ricci Dark Energy: Early-Time Inflation and Late-Time Acceleration and RGUP Corrected Observables}}

\author[1]{Khandro K Chokyi\thanks{\href{mailto:khankalmaths2023@outlook.com}{khankalmaths2023@outlook.com}}}
\author[2,3,4]{Abdel Nasser Tawfik\thanks{\href{mailto:a.tawfik@fue.edu.eg}{a.tawfik@fue.edu.eg}}}
\author[1]{Surajit Chattopadhyay\thanks{* Corresponding author: 
\href{mailto:surajitchatto@outlook.com}{surajitchatto@outlook.com}, 
\href{mailto:schattopadhyay1@kol.amity.edu}{schattopadhyay1@kol.amity.edu}}}

\affil[1]{\small Department of Mathematics, Amity University Kolkata, Kolkata 700135, India}
\affil[2]{\small Ahram Canadian University, Faculty of Engineering, Basic Science Department, 12573 Giza, Egypt}
\affil[3]{\textcolor{black}{Physics Department, Faculty of Science, Islamic University of Madinah, 13518 Madinah, KSA}}
\affil[4] {The Egyptian Center for Theoretical Physics (ECTP), Giza, Egypt}

\date{\today}

\begin{document}

\maketitle

\begin{abstract}
This study investigates a cosmological scenario within the $f(Q,L_m)$ gravity framework to explore whether one geometric model can simultaneously describe the early and late-time accelerated epochs. Motivated by the recently proposed $f(Q,L_m)$ gravity framework by Hazarika \textit{et. al.} Phys. Dark Universe 50 (2025) 102092,we adopt a minimal polynomial form, $f(Q,L_m) = -Q + \alpha Q^2 + 2L_m + \beta QL_m$, and the late-time dynamics are reconstructed by introducing Holographic Ricci Dark Energy (HRDE) as an effective fluid. The resulting background evolution demonstrates smooth accelerated expansion, stable Hubble parameter behavior, and an effective equation of state that approaches the de Sitter regime. Bayesian analysis utilizing Pantheon supernovae, cosmic chronometer, and DESI BAO data reveals that the matter–geometry coupling parameter $\beta$ is weakly constrained and remains consistent with the $\Lambda$CDM limit. In the high-curvature regime characteristic of the early Universe, the quadratic non-metricity term $\alpha Q^2$ dominates the dynamics, resulting in a Starobinsky-like inflationary phase driven solely by geometric effects with predicted $n_s$ and $r$ values consistent with PLANCK 2018 observations. Furthermore, quantum-gravity-inspired corrections are examined through a Relativistic Generalized Uncertainty Principle (RGUP), implemented as a momentum-dependent deformation of the effective spacetime metric. These corrections maintain the geometric inflationary background while introducing minor perturbative shifts in higher-order inflationary observables, specifically the running of the spectral index. Overall, these findings indicate that the $f(Q,L_m)$ framework offers a dynamically consistent geometric model in which early and late cosmic acceleration arise from distinct curvature regimes, with RGUP effects causing sub-leading modifications.
\end{abstract}

\tableofcontents
\section{Introduction}
Ever since Einstein introduced his revolutionary theory of general relativity(GR) \citep{einstein1922general} in 1915, it has remained unaltered for over a hundred years and is very fundamental to the study of cosmology and astrophysics. It is the currently accepted gravitational theory and a large number of tests and observations support it \citep{asmodelle2017tests, blair2021quest, krishnendu2021testing, turyshev2008experimental, psaltis2019testing}. The theory was so revolutionary that our view of gravity was completely transformed, positing that gravity is an intrinsic property of spacetime, rooted in Riemannian geometry rather than a conventional force.

The Friedmann equations \citep{uzan2001dynamical} are obtained by solving the \textcolor{black}{GR} equations while considering a homogeneous and isotropic spacetime described by the Friedmann–Lemaître–Robertson–Walker (FLRW) metric \cite{friedman1922krummung, lemaitre1931homogeneous, robertson1935kinematics, walker1937milne}. When GR is applied to cosmological data, it encounters well-known conceptual and observational difficulties despite its empirical success \citep{shapiro1972testing, eddington1938problem, wheeler1962problems}. 
The early Universe presents one of the first of these difficulties. The observed large-scale homogeneity, isotropy, and near-flatness of the Universe cannot be explained by classical GR along with standard matter and radiation. To rectify these shortcomings, the notion of cosmological inflation was independently proposed by  Guth\citep{guth1981inflationary}, Linde \citep{linde1982new} and others \citep{albrecht1982cosmology, starobinsky1982dynamics} in the early 1980s, according to which, the early Universe had a brief phase of fast, exponential growth, usually lasting between $10^{-36}$ and $10^{-32}$ seconds following the Big Bang. In order to smooth out any initial inhomogeneities and anisotropies, the scale factor $a(t)$ rose by many orders of magnitude over this epoch. Inflation changed our view of the early Universe, much like \textcolor{black}{GR} had changed our knowledge of gravity. This inflationary paradigm was groundbreaking in and of itself because while it provided a solution to the horizon \citep{bolotin2013thousand} and flatness \citep{helbig2012there} problem of standard cosmology, inflation also served as a mechanism that explained the observed large-scale structure through primordial density perturbations via quantum fluctuations of the inflaton field. 

Cosmic expansion was predicted to slow down due to gravity for a few decades after inflation. However, the Universe is currently going through a second phase of accelerated expansion, according to observations made by the High-z Supernova Search Team \citep{filippenko1998results} and the Supernova Cosmology Project \citep{perlmutter1999measurements}, two separate organisations which claimed in 1998 that distant Type Ia supernovae seemed fainter than anticipated. The only coherent answer was that the expansion of the \textcolor{black}{U}niverse is speeding rather than slowing down. By adding a modest, positive cosmological constant $\Lambda$ to Einstein's field equations, the simplest theoretical explanation for this accelerated expansion is the so-called $\Lambda$CDM model \citep{turner1997case}, which is still the accepted standard model of cosmology today. Despite its phenomenological success, $\Lambda$CDM model has serious theoretical problems \citep{turner2025everyone}, particularly the cosmic coincidence problem and the fine-tuning problem.

A fundamental question is raised by the existence of two separate epochs of accelerated expansion: inflation in the early Universe and acceleration driven by dark energy in later times. Can these phenomena be attributed to the geometric aspects of gravity itself? Modified theories of gravity \citep{shankaranarayanan2022modified}, which credit the gravitational sector rather than ad hoc matter components for the cosmic acceleration, have been developed in response to this question.
The Einstein–Hilbert Lagrangian has been generalised to a function of geometric invariants in a broad class of modified gravity theories as a result of these considerations. Among them, $f(R)$ gravity \citep{kerner1982cosmology, capozziello2008cosmography, sotiriou2010f, capozziello2012dark, das2023modified, starobinsky2007disappearing, de2006f, mustafa2020stable, santos2007energy, odintsov2025confronting, odintsov2025power} is the oldest and most studied. It modifies the field equations while maintaining general covariance by extending the action's Ricci scalar $R$ to an arbitrary function $f(R)$. The teleparallel equivalent of general relativity (TEGR) \citep{unzicker2005translationeinsteinsattemptunified} gave rise to alternative formulations, such as $f(T)$ gravity \citep{capozziello2011cosmography,myrzakulov2011accelerating,cai2016f,paliathanasis2016cosmological, li2011degrees, li2011large, krvsvsak2016covariant, duchaniya2022dynamical}, in which torsion, not curvature, is responsible for gravity. A third, dynamically rich method has just been proposed: the symmetric teleparallel formulation \citep{nester1999symmetricteleparallelgeneralrelativity}, where gravity is fully given by the non-metricity scalar $Q$ and both curvature and torsion vanish, i.e. the $f(Q)$ gravity. Furthermore, adding an explicit link between matter and geometry in the Lagrangian is an even more expansive and physically driven extension. The $f(Q, L_m)$ gravity theory \citep{hazarika2024f} follows from this, where $L_m$ is the matter Lagrangian density. The energy–momentum tensor is essentially non-conserved due to the direct energy and momentum transfer between the matter and geometric sectors made possible by the addition of the mixed term $Q, L_m$. This coupling mechanism enhances cosmological dynamics and can explain transitions between multiple cosmic epochs within a single theoretical framework. Thus, from the inflationary epoch to the current acceleration, the $f(Q, L_m)$ framework provides an excellent setting for examining the unified behavior of cosmic history under a single geometric base. By properly choosing the functional form of $f(Q, L_m)$, one can recover \textcolor{black}{GR}, $f(Q)$ gravity, or other feasible subcases as limiting scenarios. Furthermore, by using a number of dark energy models as effective fluids, this theory offers sufficient flexibility to close the gap between phenomenological and geometric approaches to cosmic acceleration. One can find various studies considering $f(Q,L_m)$ done in the literature\citep{MYRZAKULOV2025139506, MYRZAKULOV2024164, MYRZAKULOV2024101614, kshirsagar2026cosmologicaldynamicshyperbolicevolution, ah, MYRZAKULOV2025101829, samaddar2025novel, shiravand2025cosmologicalinflationfqmathcallmgravity, samaddar2025observationalsignaturesscalarfield}.

Concurrently, methods inspired by quantum gravity have become more and more significant in the development of cosmological models. In particular, the holographic dark energy (HDE) model, in which the dark energy density is related to a cosmological length scale, was proposed by M. Li \citep{LI20041} as a result of the holographic principle, which relates ultraviolet and infrared cutoffs in quantum field theory. Among these, the Holographic Ricci Dark Energy (HRDE) model \citep{PhysRevD.79.043511}, in which the Ricci scalar determines the infrared cutoff, naturally incorporates local spacetime dynamics through the Hubble parameter and its derivatives while avoiding the causality problems and the coincidence problem of dark energy. Thus, HRDE models have been widely investigated \citep{pasqua2013reconstruction, pasqua2025generalizedholographicriccidark, hrde, gargee, Albarran_2015, Das2015, khandro, Singh2019, particles7030051} as potential solutions to other problems in modern cosmology.

Discussions arising from string theory \citep{schwarz1999string, yoneya1989interpretation, AMATI198941, GROSS1988407, Veneziano}, black-hole thermodynamics \citep{bekenstein1980black}, and doubly special relativity \citep{amelino2002doubly, Kowalski-Glikman2005, amelino2002} point to a breakdown of the standard Heisenberg uncertainty relation \citep{busch2007heisenberg} and the appearance of a minimum measurable length \citep{Hossenfelder2013} or a maximum observable momentum at energy scales \citep{petruzziello2021generalized} that are close to the Planck mass. 
Several Generalized Uncertainty Principle (GUP) proposals \citep{tawfika, raghavi2024unified, tawfikb, ali2013effect, MAGGIORE199365}, which alter canonical commutation relations and result in corrections to quantum mechanics and field theory that are suppressed by powers of the Planck length or energy scale, encapsulate this concept. One can find a review and phenomenological study on multiple GUP formulations and its connections from black-hole thermodynamics to cosmology undertaken by A.N. Tawfik and A. Diab \citep{tawfik2014generalized} in the literature. Momentum-dependent corrections to the effective spacetime description are naturally produced by a relativistic or "renormalized" extension of these concepts (often referred to as RGUP or Relativistic GUP), which encourages the deformation to a Lorentz-consistent form. This concept was first introduced by Todorinov et. al. in their work \citep{todorinov2019relativistic}. The GUP/RGUP modification can be mapped to either modified dispersion relations for propagating fields or an effective deformation of the background metric in many concrete implementations. In practice, this is accomplished by (i) modifying the field commutators and obtaining corrected stress-energy tensors, or (ii) by determining an effective metric deformation that replicates the same corrected energy-momentum relations. Both approaches result in adjustments to normalization, sound speed, and the inflaton kinetic term, which alter slow-roll parameters and inflationary observables like the scalar spectral index $\eta_s$, its runninf $\alpha_s$ and the tensor-to-scalar ratio $r$. These corrections enter the scalar and fermionic field equations and alter the stress-energy components used in cosmological Friedmann equations, as demonstrated by recent explicit RGUP treatments \citep{bhandari2025rgup}. Since RGUP corrections depend on momentum or energy, they selectively act on modes with trans-Planckian origins which result in small but model-dependent fingerprints on the CMB’s primordial spectra \citep{danielsson2002note}. Therefore, there are several studies in the literature that use inflationary data to try to constrain the GUP and RGUP deformation parameters \citep{mughal2021relativistic, corda2025quantum, amelino2013quantum, heidarian2025alpha, bhandari2024gup}. Early on, studies showed that having a minimal length can shift the predicted \textcolor{black}{tensor-to-scalar ratio ($r$) and the scalar spectral index ($n_s$)} values, which changes how we read results from experiments like BICEP2 \citep{ade2014bicep2} and Planck \citep{planck2006scientific}. More recently, studies have been incorporating GUP and RGUP corrections into the slow-roll equations or the dispersion relations and have quantified the shift in observables \citep{tawfik2013effects}. They have found that while the effects are usually small, they’re not completely insignificant and with the next generation of CMB experiments—like the Stage-IV missions \citep{zhang2022transitioning}—there’s a potential to spot these subtle effects.

\textcolor{black}{In the present work, the main new point of our study is that we use the $f(Q,L_m)$ gravity framework where matter and geometry are directly coupled through the $Q L_m$ term. In many standard $f(Q)$ or $f(R)$ models, inflation and late-time acceleration are usually explained separately or by adding extra scalar fields. But in our model, both early-time inflation and late-time acceleration come naturally from different curvature limits of the same gravitational theory. When the curvature is very high, the quadratic non-metricity term becomes important and gives a Starobinsky-like inflation. When the curvature becomes small at late times, the matter–geometry coupling term controls the accelerated expansion. In this way, the model gives a single geometric theory to explain both phases of cosmic acceleration without adding extra artificial components.}

\textcolor{black}{The paper is organized as follows. In Section \ref{sec2}, the basic formalism of the $f(Q, L_m)$ gravity theory is presented  and  the general field equations are derived. In Section \ref{sec3} we apply this formalism to the flat FLRW \textcolor{black}{U}niverse and obtain the modified Friedmann equations. Section \ref{sec4} is devoted to the reconstruction of the cosmological model by considering Holographic Ricci Dark Energy and discussing the late-time behavior of the Universe. In Section \ref{sec5}, we perform Bayesian analysis using supernovae, cosmic chronometer and BAO data to constrain the model parameters. Section \ref{sec6} studies the inflationary dynamics in the high-curvature limit and compares the predictions with Planck 2018 results. In Section \ref{sec7}, we introduce RGUP corrections and derive the modified effective energy density and pressure. Section \ref{sec8} presents the RGUP-corrected inflationary results and their observational comparison. Finally, in Section \ref{sec9}, we summarize our findings and give concluding remarks.
}


\section{Basic formalism of \texorpdfstring{$f(Q,L_m)$}{f(Q,Lm)} gravity theory}\label{sec2}

For the purpose of providing the theoretical background for our subsequent cosmological reconstruction and analysis, we will first discuss the basic formalism of $f(Q,L_m)$ gravity. As mentioned earlier, this theory is based on the symmetric teleparallel geometry, in which gravity is described by the non-metricity of spacetime, and the curvature and torsion of spacetime are set to zero. The addition of the matter Lagrangian $L_m$ results in a direct coupling between geometry and matter, thus enabling a non-conserved energy-momentum tensor. Below, we will describe the action, geometric variables, and field equations that will be used in our analysis.
In this scenario, the gravitational action adopts the form \citep{hazarika2024f}
\begin{equation}
 S=\int f(Q,L_m)\sqrt{-g}d^4x   
\end{equation}
Here, $\sqrt{-g}$ is the determinant of the metric tensor that guarantees covariance under coordinate transformation. 
In the context of the Weyl-Cartan connection $Y^{\alpha}_{\mu\nu}$, the non-metricity tensor $Q^{\alpha}_{\mu\nu}$ is defined as the covariant derivative of the metric tensor and is mathematically expressed as given
\begin{equation}\label{g1}
Q_{\alpha\mu\nu}=\nabla_{\alpha}g_{\mu\nu}=\delta_{\alpha}g_{\mu\nu}-Y^{\beta}_{\alpha\mu}g_{\beta\nu}-Y^{\beta}_{\alpha\nu}g_{\mu\beta}
\end{equation}
In addition, the superpotential $P^{\alpha}_{\mu\nu}$ is introduced as a quantity that plays the role of the conjugate to the non-metricity tensor. It is defined as \citep{xu2019f},
\begin{equation}\label{g2}
P^{\alpha}_{\mu\nu}=-\frac{1}{2}L^{\alpha}_{\mu\nu}+\frac{1}{4}(Q^{\alpha}-\tilde{Q}^{\alpha})g_{\mu\nu}-\frac{1}{4}\delta^{\alpha}_{(\mu} Q_{\nu)}
\end{equation}
where $ Q^\alpha = Q^\alpha_{\ \mu}{}^\mu$ and $\tilde{Q}^\alpha = Q^\mu_{\ \mu}{}^\alpha$ are known as the non-metricity vectors. Finally, the non-metricity scalar $Q$ is obtained by contracting the superpotential with the non-metricity tensor
\begin{equation}\label{g3}
Q = - Q^\lambda_{\mu\nu} P^\lambda_{\mu\nu}.
\end{equation}
By the use of the varying action principle, one can obtain, 
\begin{equation}\label{b2}
    \delta S=\int [(f_Q\delta Q+f_{L_m}\delta L_m)\sqrt{-g}]d^4x
\end{equation}
where $f_{L_m}=\frac{\delta f(Q,L_m)}{\delta L_{m}}$ and $f_Q=\frac{\delta f(Q,L_m)}{\delta Q}$. Furthermore \citep{landau1975classical},
\begin{equation}\label{b3}
   T^{\mu\nu}=g_{\mu\nu}L_{m}-\frac{2\delta L_m}{\delta g^{\mu\nu}} 
\end{equation}
is the energy-momentum tensor of the matter. The variation of the determinant of the metric tensor is given by
\begin{equation}\label{b4}
    \delta\sqrt{-g}=-\frac{1}{2}\sqrt{-g}g_{\mu\nu}\delta g^{\mu\nu}
\end{equation}
Ultimately, one can derive the field equation for $f(Q,L_{m})$ gravity as
\begin{align}\label{b5}
 \frac{2}{\delta\sqrt{-g}}\nabla_{\alpha}(f_{Q}\sqrt{-g}P_{\alpha\mu\nu})&+f_{Q}(P_{\mu\alpha\beta}Q_{\nu}^{\alpha\beta}-2Q_{\mu}^{\alpha\beta}P_{\alpha\beta\nu})+\frac{1}{2}fg_{\mu\nu}\\&
 =\frac{1}{2}f_{L_m}(g_{\mu\nu}L_{m}-T_{\mu\nu})    
\end{align}
\textcolor{black}{We note that} Eq. \eqref{b5} reduces to the field equation for $f(Q)$ gravity when $f(Q,L_m)=f(Q)+2L_{m}$ as seen in \citep{wang2022static}. For a detailed description of the geometry of the modified $f(Q,L_{m})$ gravity, one can refer to \citep{hazarika2024f}. In the next section, we will discuss these general field equations within the context of a homogeneous and isotropic FLRW spacetime to obtain the cosmological dynamics relevant to the current analysis.
\section{Field equations of FLRW universe in \texorpdfstring{$f(Q,L_{m})$}{f(Q,Lm)} gravity}\label{sec3}
As we have discussed the general equations of $f(Q,L_m)$ gravity in the previous section, we now encode this formalism into a homogeneous and isotropic universe. This is a crucial step in linking the underlying geometric theory to the observable evolution of the Universe. In this work, we shall focus on a spatially flat cosmology, which is consistent with the current observational constraints imposed by the cosmic microwave background radiation.
In this context, the matter content of the Universe is described by an effective perfect fluid, which enables us to describe the modified gravitational dynamics in terms of an effective energy density and pressure. This helps us to understand the modified gravitational dynamics in terms of an effective fluid contribution that arises from the matter-geometry coupling in $f(Q,L_m)$ gravity.

In our work, we have assumed the flat FLRW metric, which is given as \citep{ryden2004introduction},
\begin{equation}\label{1}
    ds^2=-dt^2+a^2(t)(dx^2+dy^2+dz^2)
\end{equation}
where $a(t)$ is the scale factor. The rate of expansion is given by the Hubble parameter $H=\frac{\dot{a}}{a}$ where the dot represents the derivative with respect to cosmic time. When we assume the universe to be filled with a perfect fluid, the energy-momentum tensor is given as
\begin{equation}\label{2}
    T_{\mu\nu}=(\rho+p)\mu_{\mu}\mu_{\nu}+pg_{\mu\nu}
\end{equation}
where $p$ is the isotropic pressure, $\rho$ is the energy density, and $\mu^{\mu}$ is the 4-velocity of the fluid with components $\mu^{\mu}=(1,0,0,0)$. The matter Lagrangian $L_m$ in theories with non-minimal matter–geometry coupling, like $f(Q,L_m)$ gravity, is not uniquely defined for a perfect fluid, and different choices (like $L_m = -\rho$ or $L_m = p$) can result in different cosmological dynamics because $L_m$ explicitly appears in the field equations. This argument has been revisited in the work of V. Faraoni \citep{faraoni2009lagrangian}. The author shows that the choice of $L_m=p$ or $L_m=\rho$ is insignificant while considering perfect fluids that are not directly coupled to gravity. This has also been shown in \citep{brown1993action}. The author further notes that in cases where the gravity couples with the fluid, the two density forms of Lagrangian are not equivalent.  Finally, the above-mentioned study concludes that since the two Lagrangian densities produce two inequivalent theories of gravity and matter, both of which are accurate, it is unclear which should be physically preferred over the other. It emphasizes that independent reasoning must be used to determine which should be selected. Bertolami et. al. \citep{bertolami2008non} noted that $L_m=-\rho$, which is produced by appending surface terms to the action $S=\int d^4x\sqrt{-g}L_m$, is an equivalent Lagrangian density for a perfect fluid (for in-depth analyses of the Lagrangian formalism for perfect fluids, \textcolor{black}{see \citep{faraoni2009lagrangian}} ). It would seem from that the decision between the two comparable Lagrangians determines whether the additional force that could replace dark matter is present or absent. This demonstrates that $L_m=p$ and $L_m=-\rho$ are not equivalent. In our work, we have chosen $L_m = p$, which is frequently employed in the literature on $f(R,L_m)$ \citep{harko2010f} and $f(Q,L_m)$ theories\citep{hazarika2024f}. This decision guarantees a consistent perfect-fluid description and prevents the non-conservation equation of the energy-momentum tensor from introducing unphysical extra force terms, which could occur with other choices like $L_m = -\rho$. Furthermore, the pressure-based Lagrangian offers a stable and physically well-motivated effective fluid interpretation for a homogeneous and isotropic cosmological background, which has been widely used in matter–geometry coupling models. Consequently, the assumption $L_m = p$ is consistent with previous research in non-minimally coupled modified gravity frameworks and permits a consistent reconstruction of cosmological dynamics.
This choice prevents the non-conservation equation of the energy-momentum tensor from introducing extra force terms, which could occur with other choices like $L_m = -\rho$. Furthermore, the assumption $L_m = p$ is consistent with previous research \citep{hazarika2024f} in non-minimally coupled modified gravity frameworks and permits a consistent reconstruction of cosmological dynamics.
Lagrangian.\\
The two generalized Friedmann equations for the dynamics of FLRW universe in the framework of $f(Q,L_{m})$ gravity are obtained as \citep{myrzakulov2025constraining},
\begin{equation}\label{3}
    3H^2=\frac{1}{4f_{Q}}[f-f_{L_{m}}(\rho+L_m)],
\end{equation}
\begin{equation}\label{4}
   \dot{H}+3H^2+\frac{\dot{f_{Q}}}{f_{Q}}H=\frac{1}{4f_{Q}}[f+f_{L_{m}}(p-L_m)] 
\end{equation}
\textcolor{black}{Using Eq. \eqref{3}}, we can rewrite Eq. \eqref{4} as
\begin{equation}\label{5}
  2\dot{H}+3H^2=\frac{1}{4f_{Q}}[f+f_{L_m}(\rho+2p-L_{m})]-2\frac{\dot{f_{Q}}}{f_{Q}}H
\end{equation}
Thus, we can write,
\textcolor{black}{\begin{align}\label{6}
&3H^{2}=\rho_{\mathrm{eff}}~~~ \mathrm{and}\\ 
&2\dot{H}+3H^2=-p_{\mathrm{eff}}
\end{align}}
Here, \textcolor{black} {$\rho_{\mathrm{eff}}$} is the effective energy density and 
\textcolor{black}{\begin{equation}\label{7}
 \rho_{\mathrm{eff}}=\frac{1}{4f_{Q}}[f-f_{L_{m}}(\rho+L_m)]   
\end{equation}}
Similarly, the effective pressure can be defined as
\textcolor{black}{\begin{equation}\label{8}
 p_{\mathrm{eff}}=2\frac{\dot{f_{Q}}}{f_{Q}}H-\frac{1}{4f_{Q}}[f+f_{L_m}(\rho+2p-L_{m})]   
\end{equation}}
The set of modified Friedmann equations \eqref{3}-\eqref{8} above provides the dynamical system that describes the evolution of the cosmos in the $f(Q, Lm)$ theory. All the deviations from GR are then captured by the modified energy density, \textcolor{black}{$\rho_{\mathrm{eff}}$}, and pressure, \textcolor{black}{$p_{\mathrm{eff}}$}. Thus, in the next section, we will discuss the reconstruction of our cosmological model.
\section{Cosmological model of \texorpdfstring{$f(Q,L_{m})$}{f(Q,Lm)} gravity with Holographic Ricci Dark Energy}\label{sec4}
In this section, we have explained in detail the reconstruction of our specific cosmological model through which we can study the dynamics of the Universe. Our strategy consists of choosing a physically inspired functional form of $f(Q, L_m)$ and then recovering the corresponding cosmological evolution by considering an effective dark energy component of holographic type. This strategy allows for the investigation of the possibility of simultaneously explaining the early-time inflationary and late-time acceleration phases of the Universe's evolution within a geometric framework.

Thus, we have considered \citep{hazarika2024f}
\begin{equation}\label{c1}
f(Q,L_m) = -Q + \alpha Q^2 + 2L_m + \beta QL_m
\end{equation}
which is the minimal but physically non-trivial generalization of the symmetric teleparallel formulation of gravity. \textcolor{black}{Each term in Eq. \eqref{c1} has a physical interpretation
\begin{itemize}
    \item The linear term, $-Q + 2L_m$, is a replica of the usual GR limit and ensures the theory simplifies to GR at $\alpha = \beta = 0$.
    \item The quadratic term $\alpha Q^2$ reflects the leading-order $Q$ correction, commonly justified by quantum or high-energy origins, and is important for the description of early-universe dynamics like inflation or non-singular bounces.
    \item The addition of the term $\beta QL_m$ allows for a non-minimal coupling between matter and geometry, creating possible energy transfer between these sectors and a violation of the usual conservation law of the energy–momentum tensor. Such a matching makes it possible to observe late-time cosmic acceleration in a controlled way. 
\end{itemize}} The constants $\alpha$ and $\beta$ are used as small perturbative constants in order for the model to be close to the GR regime and in agreement with observational limits. Furthermore, the polynomial form of Eq.~(\ref{c1}) keeps the field equations analytically solvable, allowing for an easy inspection of both early- and late-time cosmological evolution.Thus, \begin{equation}\label{9}
 f_Q=\frac{df}{dQ}=-1+2\alpha Q+\beta L_m   
\end{equation} 
and 
\begin{equation}\label{10}
f_{L_m} =\frac{df}{dL_m}=\beta Q+2  
\end{equation}
In order to describe the dark energy sector in our reconstructed model, we add Holographic Ricci Dark Energy (HRDE) component. Because of its intriguing theoretical underpinnings and phenomenological congruence with observations, the HRDE model has drawn the most attention among the other holographic dark energy ideas. The model's foundation is the holographic principle, which links the effective quantum field theory's ultraviolet (UV) and infrared (IR) cutoffs. It implies that the total energy in an area of size $L$ shouldn't be greater than the mass of a black hole of the same size. The HRDE model is well-suited to symmetric teleparallel gravity, since the Ricci scalar, and therefore the HRDE density, can be written in terms of the Hubble parameter and its derivatives. These parameters are directly linked to the non-metricity scalar $Q = 6H^2$ in the FLRW background. Unlike the horizon-scale holographic models, the HRDE model uses a local infrared cutoff based on the curvature of spacetime, thus avoiding any possible nonlocal problems with causality that could emerge in theories with direct matter-geometry coupling. From a geometric perspective, the HRDE model provides a natural bridge between energy scales in quantum gravity and modified gravity theories. Its $H$ and $\dot{H}$ dependence allows the dark energy density to dynamically adapt to changes in the rate of expansion, thus allowing it to drive both the inflationary era and the accelerating era of the universe in a single scenario. This dynamical behavior is particularly useful in $f(Q,L_m)$ gravity, since the difference from General Relativity is already encoded through higher-order non-metricity corrections and matter-geometry couplings.

Thus, the HRDE density is given by
\begin{equation}\label{c5}
\rho_{\mathrm{HRDE}} = 3 c^2 (2H^2 + \dot{H}), 
\end{equation}
where $c^2$ is a dimensionless constant describing the strength of the holographic contribution.
We know that the conservation equation is,
\begin{equation}\label{11}
    \dot{\rho}_{\mathrm{HRDE}}+3H(\rho_{\mathrm{HRDE}}+p_{\mathrm{HRDE}})=0
\end{equation}
can be used to derive the pressure related to HRDE as
\begin{equation}\label{12}
p_{\mathrm{HRDE}}=-\frac{\dot{\rho}_{\mathrm{HRDE}}}{3H}-\rho_{\mathrm{HRDE}}
\end{equation}
In our work, as the HRDE part is represented as an effective fluid, the pressure can be equated to the matter Lagrangian density $L_m$ in the $f(Q, L_m)$ formalism.Thus, using \textcolor{black}{Eq. \eqref{12}}, we get
\begin{equation}\label{13}
    \rho_{\mathrm{HRDE}}+L_m=-\frac{\dot{\rho}_{\mathrm{HRDE}}}{3H}
\end{equation}
 From \textcolor{black}{Eq. \eqref{c5}}, we obtain
 \begin{equation}\label{14}
    \dot{\rho}_{\mathrm{HRDE}}=3c^2(\ddot{H}+4H\dot{H}) 
 \end{equation}
which was then substituted in \textcolor{black}{Eq. \eqref{13}} to give
\begin{equation}\label{15}
\rho_{\mathrm{HRDE}}+L_m=-\frac{c^2\ddot{H}}{H}-4c^2\dot{H}
\end{equation}
\textcolor{black}{Substituting Eqns. \eqref{c1}, \eqref{9}, \eqref{11} and Eq. \eqref{15} into the modified Friedmann equation \eqref{3} and taking $Q=6 H^2$,  the equation reduces to}
\begin{equation}\label{16}
    12 H^2=\frac{-6H^2+36\alpha H^4-36\beta c^2H^4-18\beta c^2H^2\dot{H}-6c^2\dot{H}-12c^2 H^2}{-1+12\alpha H^2-\beta c^2\frac{\ddot{H}}{H}-7\beta c^2\dot{H}-6 \beta c^2 H^2}
\end{equation}
which then finally gives us the second derivative of the Hubble parameter as
\begin{equation}\label{17}
    \ddot{H}=\frac{6H^4(3\alpha-\beta c^2)-c^2\dot{H}(11\beta H^2-1)-H^2(1-2c^2)}{2\beta c^2 H}
\end{equation}

Equation \eqref{17} is the main \textcolor{black}{expression} that describes the evolution of the universe in our reconstructed $f(Q, Lm)$-HRDE model. As seen above, it is a second-order nonlinear differential equation for the Hubble parameter, where the effects of higher-order non-metricity corrections and matter-geometry coupling are explicitly incorporated through the parameters $\alpha$ and $\beta$. The presence of both $\dot{H}$ and $\ddot{H}$ terms captures the inherently dynamical nature of the holographic Ricci dark energy component, whose energy density is a function of spacetime curvature and its evolution.

Physically, Eqn. \eqref{17} can admit different evolution phases. During the early \textcolor{black}{U}niverse, when the Hubble parameter takes large values, the quadratic non-metricity term proportional to $\alpha$ dominates, leading to an accelerated evolution phase consistent with the inflationary era. During the late \textcolor{black}{U}niverse, when the curvature scale becomes smaller, the terms proportional to the matter-geometry coupling parameter $\beta$ and the HRDE component become more important, allowing a transition from a decelerated to an accelerated evolution phase.

\begin{figure}[htbp]
    \centering
    \includegraphics[width=\linewidth]{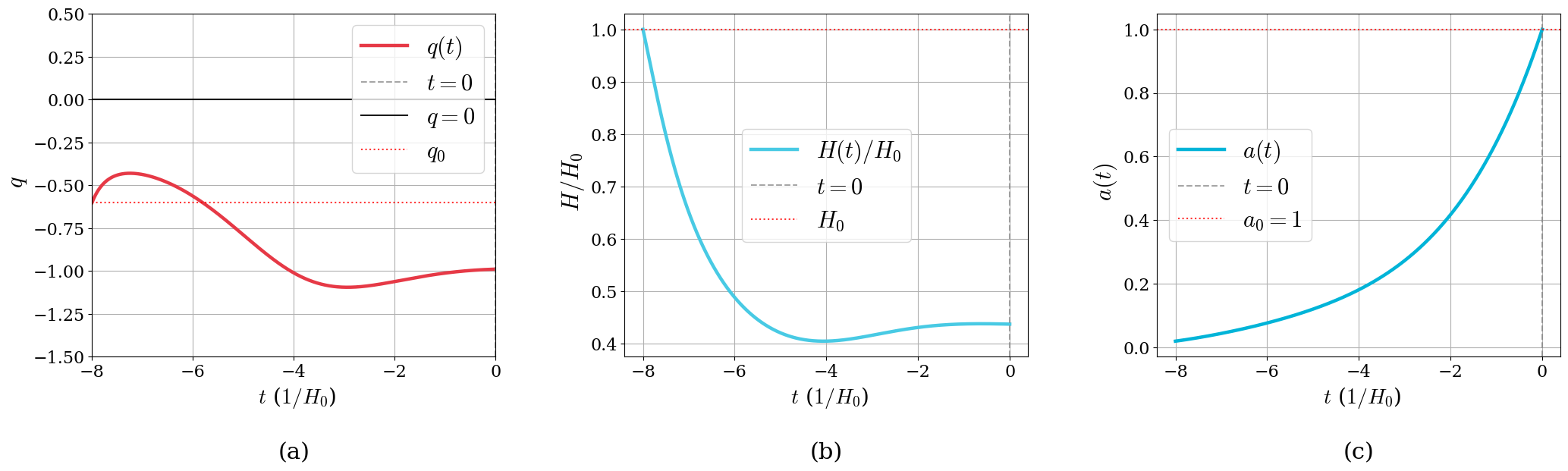}
    \caption{Time evolution of key cosmological quantities in the reconstructed $f(Q,L_m)$ model.}
    \label{f1}
\end{figure}

Because of its highly nonlinear nature, Eqn.\eqref{17} cannot be solved analytically and is instead solved numerically for the appropriate initial conditions. To this end, the second-order equation is transformed into a set of first-order differential equations, which are then solved backwards in time, starting from the present day. The present-day values of the Hubble parameter and deceleration parameter are set to $H(t_0) = H_0$ and $q(t_0) = q_0$, respectively. 
The reconstruction method allows us to follow the cosmological evolution backwards in time from the recent universe and investigate the onset of accelerated expansion due to the joint effect of non-metricity corrections and holographic dark energy. From the numerical solution, we obtain the scale factor, deceleration parameter, and equation of state, which are presented in Fig. \ref{f1}.

As seen in panel a of Fig. \ref{f1}, the deceleration parameter $q(t)$ remains negative for the entire evolution that is shown, which indicates that the Universe is in an accelerated expansion phase during the time interval considered. At the early epochs, $q(t)$ is moderately negative, indicating a mild acceleration phase, and approaches values close to $-1$ later, which is characteristic of a dark energy-dominated era. Since $q(t)$ does not change sign during the time interval shown, the acceleration era begins before the earliest time included in the numerical integration.

As for the scale factor $a(t)$, it increases smoothly and steadily from the past until the present day, with the normalization condition $a(t_0) = 1$ imposed at $t = 0$ (see panel b of Fig. \ref{f1}). There is no turning point and no contraction, which indicates that this solution describes a standard expanding universe, but not a bouncing or cyclic one. At early times, $a(t)$ increases slowly, which indicates a weakly accelerated or transitional phase. As time passes, this increase accelerates, which indicates a clear late-time acceleration due to the action of non-metricity corrections and the holographic dark energy component.

Furthermore, from panel c of Fig. \ref{f1}, one can observe that the Hubble parameter $H(t)$ decreases monotonically from higher values in the past to nearly constant values at late times. This behavior indicates the decreasing importance of matter contributions and the increasing importance of geometric and holographic corrections. The nearly flat behavior of $H(t)$ around the present epoch indicates a quasi-de Sitter phase, which is consistent with the observation of late-time cosmic acceleration. Importantly, H(t) varies smoothly without any sudden changes or instabilities, which reinforces the dynamical consistency of the reconstructed solution.

\begin{figure}[htbp]
    \centering
    \includegraphics[width=0.8\linewidth]{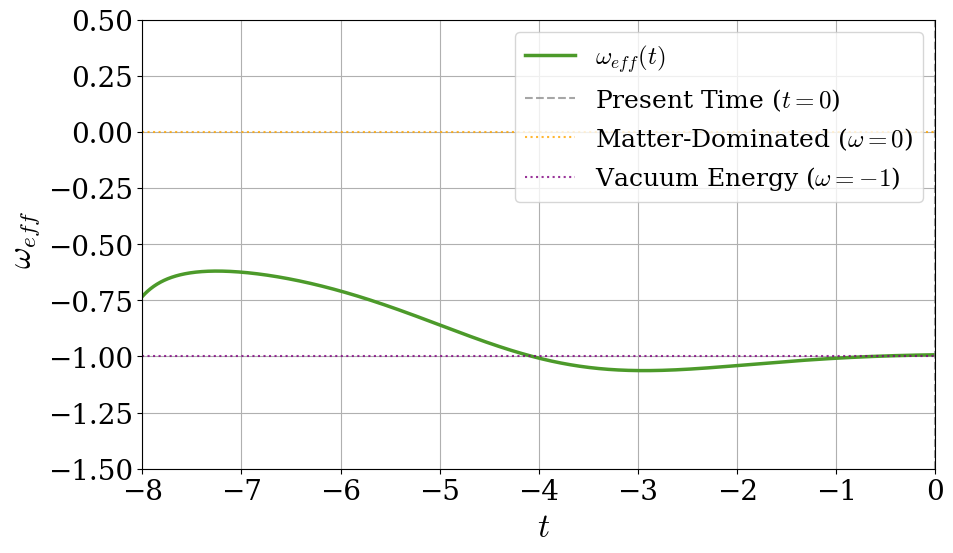}
    \caption{The behaviour of EoS parameter with respect to cosmic time $t$.}
    \label{f2}
\end{figure}

The evolution of the effective equation of state(EoS) parameter, \textcolor{black}{$\omega_{\mathrm{eff}}$}, is shown in \textcolor{black}{Fig}. \ref{f2}. Initially, in the early universe, \textcolor{black}{$\omega_{\mathrm{eff}}$} remains above -1, i.e. in a quintessence-dominated era. With the passage of time, it dips down and crosses the phantom divide, entering a phantom-dominated era. Later, \textcolor{black}{$\omega_{\mathrm{eff}}$} approaches -1 from below, indicating the onset of an effective dark energy-dominated era, entering a de Sitter-dominated phase.
This continuous evolution indicates that the matter-geometry coupling in the \textcolor{black}{$f(Q,L_m)$} model can lead to a controlled phantom era without driving the universe towards future singularities. Along with the HRDE reconstruction, the above results indicate a viable and dynamically consistent description of the late universe. The model easily leads to accelerated expansion, a stable evolution of the Hubble parameter, and an effective equation of state parameter approaching the cosmological constant value, all due to the dynamics between non-metricity corrections and matter-geometry coupling.

\section{Bayesian Inference for the \texorpdfstring{$f(Q,L_{m})$}{f(Q,Lm)} Model}\label{sec5}
A key part of the $f(Q,L_m)$ framework is how easily it separates high-curvature and low-curvature effects.
The quadratic non-metricity correction proportional to $\alpha$ is significant only during the early epoch when the non-metricity scalar $Q \sim H^2$ is substantial, whereas the matter-geometry coupling term proportional to $\beta$ predominantly influences the expansion history at later times.
This separation of scales leads to a two-step observational strategy: late-time cosmological data are used to limit $\beta$, while inflationary observables look into the high-energy parameter $\alpha$. The underlying theoretical model is defined by Eqn. \eqref{c1} where $\beta$ encodes a non-minimal coupling between matter and geometry and $\alpha$ is responsible for high-energy corrections relevant for inflation.
$\alpha$ is fixed in the current analysis because it only affects the early Universe and has little effect on late-time observables.
Instead, as covered in the subsequent section, its constraints are derived from inflationary physics.
\subsection{Late-Time Parametrization}
We use a phenomenological parametrization for the normalized Hubble function
\begin{equation}
    E^{2}(z) = \Omega_{m}(1+z)^{3}
    + (1-\Omega_{m})\left[1+\beta\,\frac{z}{1+z}\right],
\end{equation}
Here, the parameter $\beta$ effectively measures the strength of the matter-geometry coupling at low redshifts. This ansatz is not based on the exact analytical solution to the complete $f(Q, L_m)$ background equations, which are highly nonlinear. Rather, it is used to provide an effective solution to the evolution history at low redshifts.

The motivation for for this is the need to reduce the computational cost. If the full equation is to be solved inside the MCMC likelihood code, it would be numerically expensive, especially since the equation is nonlinear. The ansatz adopted effectively captures the leading deviation from the $\Lambda$CDM model caused by the coupling parameter, $\beta$, while correctly reproducing the matter-dominated evolution at high redshift. Notably, the result is identical to the standard $\Lambda$CDM evolution history when $\beta=0$.

The parameter vector sampled in the Bayesian analysis is \begin{equation} \theta = (\beta,\, M), \end{equation} where $M$ is a nuisance parameter that represents the absolute magnitude of Type~Ia supernovae. We note that Hubble’s constant $H_0$ is not varied as a parameter. Rather, the analysis employs the dimensionless expansion rate $E(z) = \frac{H(z)}{H_0},$ and the total distance scale is incorporated into the supernova nuisance parameter $M$. In this analysis, the fiducial values of $\Omega_m$ and $H_0$ are kept constant, and only the coupling parameter $\beta$ is varied. This method isolates the effect of matter-geometry coupling on late-time expansion without degeneracies with other standard cosmological parameters.

The \texttt{emcee} affine--invariant ensemble sampler \textcolor{black}{\citep{foreman2019emcee}} with 60 walkers and 12,000 steps is used for posterior sampling \textcolor{black}{\citep{heavens2009statistical}}.
The Gelman-Rubin statistic \textcolor{black}{\citep{gelman1995bayesian}} and visual examination of walker trajectories are used to evaluate convergence.
Excellent sampling efficiency and convergence are indicated by a representative run that yields $R_{\beta} = 1.005$, \qquad $R_{M} = 1.005$, with an acceptance fraction of roughly $0.66$. \textcolor{black}{We would like to clarify that the late-time parametrization used in the Bayesian analysis is only an effective description at low redshift of the full dynamical system. The complete background equation is highly nonlinear and it becomes very difficult and time-consuming to solve it repeatedly inside the MCMC process \textcolor{black}{\citep{sharma2017markov,dunkley2005fast,brooks1998markov}}. Therefore, we have used a simpler phenomenological form which captures the main effect of the matter--geometry coupling parameter $\beta$. This form is not the exact solution of the full field equations, but only an approximate expression valid in the low-redshift region. We also note that future and more precise observational data, especially from upcoming BAO \textcolor{black}{\citep{mena2026dark}} and supernova surveys, may help to constrain $\beta$ more strongly and reduce the present degeneracy seen in the posterior results.}

\subsection{Datasets and Likelihood Construction}
Three separate late-time cosmological probes are incorporated to construct the overall likelihood.
\begin{itemize}
    \item \textbf{Pantheon Supernovae}: 1048 distance--modulus measurements spanning
    $0.01 < z < 2.3$ \citep{scolnic2018complete}. We have used their official light-curve data and constructed the likelihood with their distance-modulus uncertainties, utilising the uncertainties in a diagonal covariance approximation;
    \item \textbf{Cosmic chronometers}: direct $H(z)$ measurements from the literature \citep{stern2010cosmic, moresco2012improved, moresco2015raising, zhang2014four, simon2005constraints};
    \item \textbf{DESI 2024 BAO}: Gaussian BAO summary measurements of $D_M/r_d$ and $H(z) r_d$ at effective redshifts 
$z = 0.51, 0.70, 0.92,$ and $1.32$. We have used the publicly available DESI DR1 Gaussian likelihood files which are the results referenced in \citep{adame2025desi}.

\end{itemize}
The total likelihood is
\begin{equation}
    -2\ln\mathcal{L} 
    = \chi^{2}_{\rm SN} + \chi^{2}_{H(z)} + \chi^{2}_{\rm BAO}.
\end{equation}

Flat priors are adopted for all parameters:
\[
\beta_{\rm min} \le \beta \le \beta_{\rm max},
\qquad -20 < M < -18,
\]
with several prior ranges explored to test the robustness.

\subsection{Posterior Outcomes and Parameter Limitations}

\begin{figure}[htbp]
    \centering
    \includegraphics[width=\linewidth]{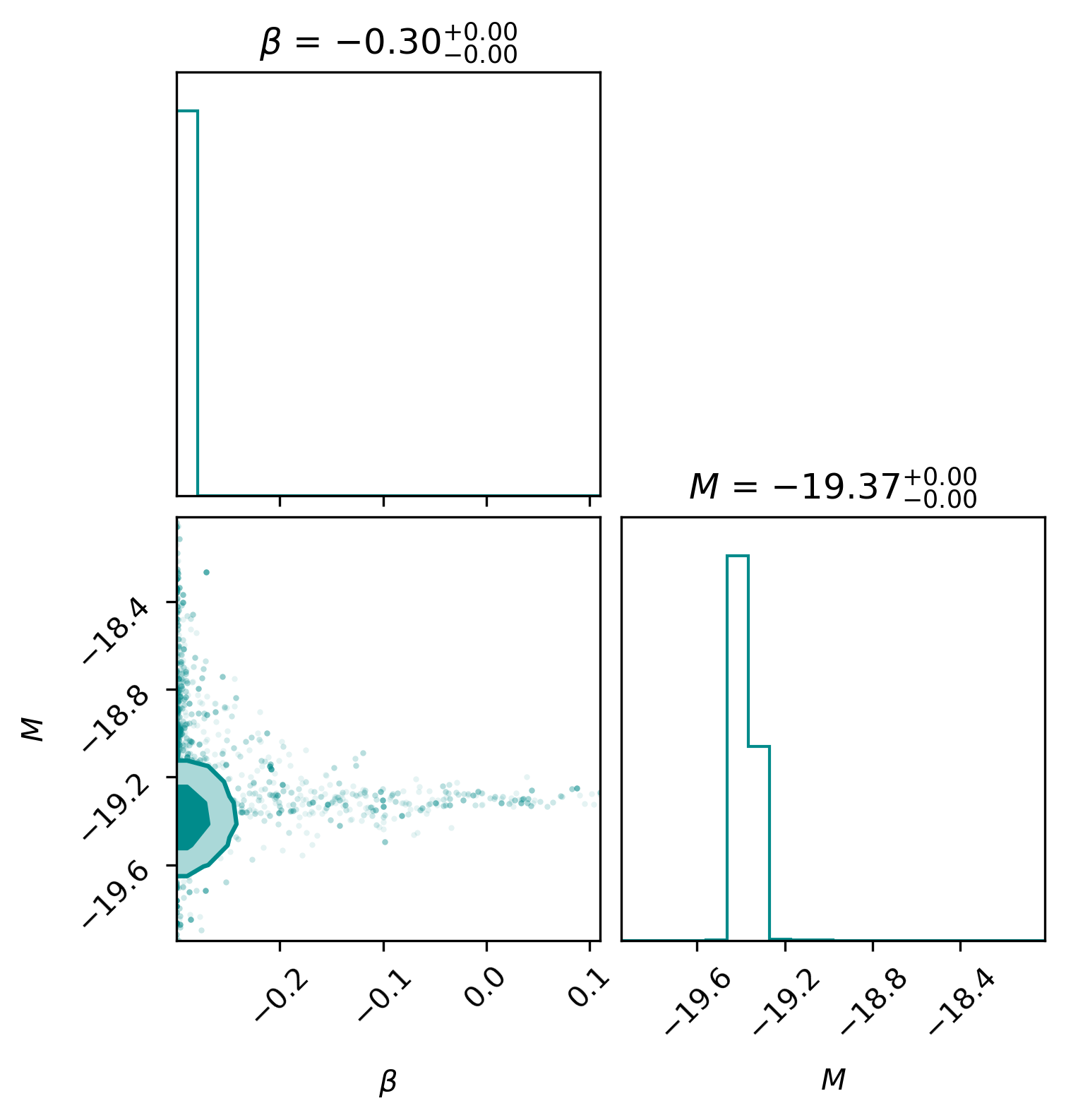}
    \caption{Corner plot for the parameters $(\beta, M)$.  
    The posterior shows strong degeneracy in $\beta$, which accumulates at the lower prior boundary, while $M$ remains sharply constrained.  
    This behaviour indicates that current late-time data do not meaningfully constrain the matter--geometry coupling.}
    \label{f3}
\end{figure}

Fig.~\ref{f3}, which shows the marginalized one-dimensional distributions and the joint confidence contours for the parameters $(\beta, M)$, summarizes the posterior distributions derived from the MCMC analysis. The posterior's qualitative behavior is unchanged by any of the tested prior widths, including $\beta \in [-1,1]$, $[-0.3,0.3]$, and $[-0.1,0.1]$.
Specifically, there is no clearly defined internal maximum in the posterior distribution for the matter-geometry coupling parameter $\beta$.
Rather than a statistically preferred non-zero value, the probability density accumulates toward the lower boundary of the prior, indicating a monotonic likelihood behavior.
Because of this, the data only offer a one-sided constraint on $\beta$, and symmetric confidence intervals should not be quoted.
Rather than a physical preference for a negative coupling, this behavior indicates an extended degeneracy between the matter--geometry coupling and the background expansion history. On the other hand, the data tightly constrain the supernova absolute magnitude $M$, which has a narrow and roughly Gaussian posterior.
Below, we give the representative summary:
\[
\beta_{\rm best} = \beta_{\rm min}, \qquad 
M = -19.37 \pm 0.004.
\]
The joint posterior further confirms that the degeneracy mainly affects the matter--geometry coupling sector by demonstrating that changes in $\beta$ have little effect on determining $M$. The limited sensitivity of existing late-time datasets to mild matter-geometry couplings should be interpreted as the reason for the collapse of the $\beta$ posterior toward the prior boundary. The data still remains entirely consistent with $\beta=0$ with respect to the minimally coupled $f(Q)$ limit and a background evolution that is similar to $\Lambda$CDM within the considered parameter space.

\section{Inflationary Dynamics in the High-Curvature Limit of \texorpdfstring{$f(Q,L_m)$}{f(Q,Lm)} Gravity}\label{sec6}
As we have already investigated the viability of the $f(Q,L_m)$ model in the late-time epochs, we now look at the dynamics of the model in the early times. One of the advantages of the model is that there is a natural separation in the low-curvature limit and high-curvature limit, hence the acceleration in the late-time and early-time inflation have a common gravitational action.

In the early Universe, when the curvature is very large, the non-metricity scalar takes the form $Q\approx 6 H^2$, which is much larger than 1. In this limit, the action is dominated by the quadratic term in the Eqn. \eqref{c1}.
compared to the linear term in $Q$ and the matter-geometry interaction term. Since the matter fields are diluted during inflation, $L_m\approx0$, and the $\beta Q L_m$ term becomes irrelevant. Therefore, the inflationary part of the action simplifies to:
\begin{equation}
    f(Q) \simeq -Q + \alpha Q^{2},
\end{equation}
This expression corresponds to the Starobinsky model $R + R^{2}$ model \citep{starobinsky1980new} where $R \rightarrow Q$. This is due to our specific minimal polynomial choice of $f(Q, L_m)$ and is not a generic characteristic of all modified gravity theories.
The evolution can be mapped to an effective scalar degree of freedom with a plateau-like potential using the standard scalar-tensor correspondence of $R+R^2$-type theories. Rather than being a fundamental inflaton field, this scalar should be viewed as a geometric auxiliary representation of the modified gravity sector. 
In the high-curvature limit, the scalar spectral index and the tensor-to-scalar ratio take the universal values of the Starobinsky model \citep{starobinsky1980new}:
\begin{align}
    n_s &\simeq 1 - \frac{2}{N}, \label{eq:infl_ns}\\[4pt]
    r   &\simeq \frac{12}{N^2}, \label{eq:infl_r}
\end{align},
where $N$ is the number of e-folds between the horizon exit of the pivot mode $k = 0.002\,\mathrm{Mpc}^{-1}$ and the end of inflation \citep{melcher2025cosmological}.

\subsection{Numerical Evaluation of Inflationary Observables}

To compare the model with observations, we calculate $(n_s, r)$ for typical e-fold numbers:
\[
N = 50,\quad 55,\quad 60,
\]
which corresponds to possible reheating in agreement with late-time cosmology. The calculated values are summarized in the Tab. \ref{tab1}

\begin{table}[h!]
\centering
\caption{Predicted inflationary observables for the high-curvature limit of the $f(Q,L_m)$ model.}
\label{tab1}
\begin{tabular}{c c c}
\toprule
$N_*$ & $n_s$ & $r$ \\
\midrule
50 & 0.96045 & $4.43 \times 10^{-3}$ \\
55 & 0.96403 & $3.69 \times 10^{-3}$ \\
60 & 0.96701 & $3.11 \times 10^{-3}$ \\
\bottomrule
\end{tabular}
\end{table}

All three predictions are comfortably inside the Planck 2018 \citep{akrami2020planck} allowed region, and the tensor amplitude remains well below the observational bound $r<0.07$.

\subsection{Comparison with Planck 2018 Constraints}

\begin{figure}[h!]
    \centering
    \includegraphics[width=0.9\linewidth]{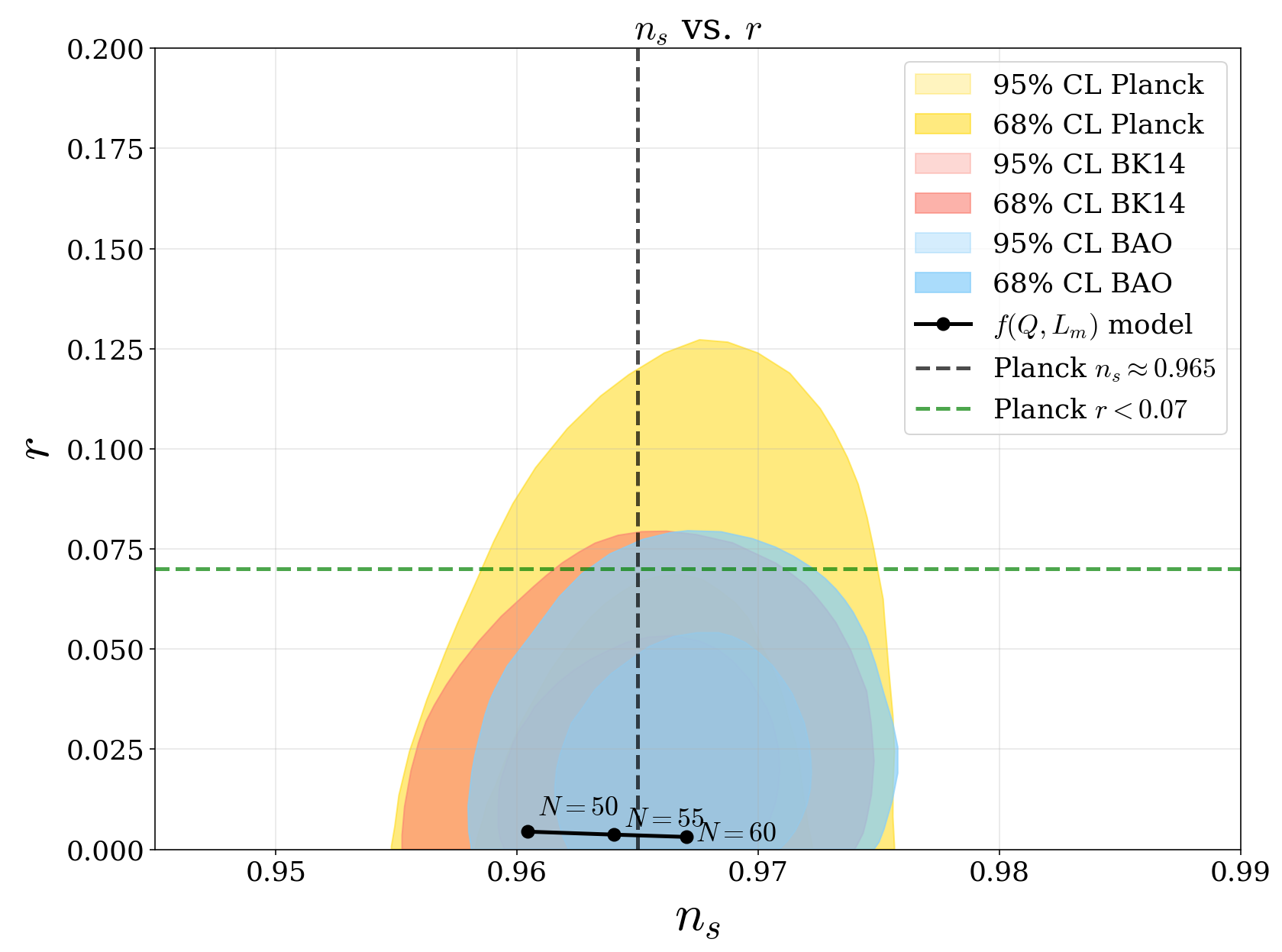}
    \caption{Constraints in the $(n_s, r)$ plane at $k = 0.002\,\mathrm{Mpc}^{-1}$ from Planck, BK14, and BAO datasets.  
    The black points denote the predictions of the high-curvature $f(Q,L_m)$ model for $N_*=50, 55, 60$.}
    \label{fig:inflation_ns_r}
\end{figure}

Figure~\ref{fig:inflation_ns_r} shows a comparison between the theoretical predictions of the $f(Q,L_m)$ model and the observational confidence contours in the $(n_s, r)$ plane obtained from \textcolor{black}{Planck 2018 \citep{akrami2020planck}, BICEP/Keck (BK14) \citep{ade2014bicep2}, and BAO data \citep{mena2026dark}.} The values of the contour coordinates are taken from publicly available tabulated likelihood contour data and are used in this work for a consistency check instead of a full likelihood analysis.

The scalar spectral index predicted by the model,
\[
0.960 \lesssim n_s \lesssim 0.967,
\]
is fully consistent with the Planck 2018 constraint $n_s = 0.9649 \pm 0.0042$.  
The predicted tensor-to-scalar ratio,
\[
r \sim 10^{-3},
\]
lies far below current detection limits while remaining observationally viable.
We would like to stress that this comparison is meant to be an observational viability test in the $(n_s,r)$ plane, as a full likelihood analysis is not within the scope of the current modified gravity model


\subsection{Physical Interpretation}
We observe that within the context of this $f(Q, Lm)$ model, inflation arises naturally from the quadratic non-metricity term without the need for an additional scalar field. The matter geometry coupling, which is brought on through the parameter $\beta$, will dictate the late-time evolution, and since this quantity is suppressed at higher curvatures, inflationary and early-time evolution, and late-time evolution, are cleanly decoupled.

Thus, the model offers:
\begin{itemize}
    \item a geometrically motivated trigger for inflation;
    \item predictions that match well with the outcomes of Planck 2018 \citep{aghanim2020planck} and BK14 \citep{ade2014bicep2} observational studies;
    \item a single framework in which inflation and late-time acceleration are consequences of different limits of the same underlying theory.
\end{itemize}
That is, the high-curvature regime of the function $f(Q, L_m)$ as demonstrated in this section, results in a strong and observationally acceptable scenario of inflation. The result is similar to the Starobinsky-like models while arising directly from symmetric teleparallel gravity's geometric properties.

\textcolor{black}{It is important to note that the high-curvature inflation phase and the low-curvature late-time phase are not two separate or disconnected parts. The same gravitational theory works for the whole cosmic evolution, but different terms become important at different curvature scales. In the early Universe, when curvature is very large, the quadratic non-metricity term becomes strong and produces inflation. As the Universe expands and curvature becomes smaller, this term slowly becomes weak and the radiation- and matter-dominated phases can take place in the usual way. At late times, the matter–geometry coupling term becomes more important and causes the accelerated expansion. In this way, all cosmic stages are connected smoothly within one single theoretical model.}

\section{Quantum–Gravity Induced Deformation of the Metric}\label{sec7}

It is theoretically well-motivated to examine whether Planck-scale quantum-gravity effects can introduce subleading corrections to this background evolution, even though the high-curvature analysis in Section \ref{sec6} shows that inflation in the current model is purely geometric and driven by the quadratic non-metricity term $\alpha Q^2$. The basic $f(Q,L_m)$ gravitational force that causes inflation is not altered in this work. Rather, we use RGUP-induced deformations of the effective spacetime metric that matter fields experience to phenomenologically incorporate quantum-gravity effects. Consequently, RGUP only functions as a perturbative quantum-geometric correction to the matter sector and inflationary observables, whereas the classical inflationary background continues to be Starobinsky-like \citep{ketov2011embedding, ketov2012inflation} and geometrically generated.

While we have seen that an acceptable inflationary scenario is obtained through the high-curvature limit of the $f(Q, L_m)$ model in Section \ref{sec6}, the very high energy scales characteristic of inflationary theory \citep{guth2005inflationary, de2015natural, ashoorioon2014reconciliation} suggest that quantum gravity corrections may not be entirely negligible. Without a fundamental theory of quantum gravity, a phenomenological approach is a useful tool to explore the possible effects of Planck-scale quantum gravity corrections \citep{jacobs2025does} to the classical theory of spacetimes.

With this observation in mind, we implement quantum gravity effects via corrections provided by Relativistic Generalized Uncertainty Principle (RGUP) \citep{todorinov2019relativistic}. The approach here does not directly modify gravity's fundamental action, but rather alters the forms of the spacetime geometry as functions of momentum, and thus affects all matter fields that propagate from the same geometry given by \(f(Q, L_m)\). This guides us to test the stability of the inflationary evolution as provided earlier and assess whether quantum gravity distortions can induce observational signatures violating standard Starobinsky attractor solution \citep{linde2014inflationary, kallosh2013superconformal}.
In the high–energy regime relevant for inflation, the background spacetime receives quantum–gravity corrections that effectively deform the metric. Following the RGUP–motivated prescription, we consider a momentum–dependent conformal deformation of the form \citep{tawfik2023born, tawfik2023quantum,tawfik2023possible}:
\begin{equation}
    \tilde g_{\mu\nu} = C(x,p)\, g_{\mu\nu},
\end{equation}
where the deformation factor $C(x,p)$ is defined as \citep{tawfik2024quantum,tawfik2025einstein, tawfik2023born1}:
\begin{equation}
    C(x,p) = \left(\phi^2 + \frac{2\kappa}{(p_0^0)^2}K^2\right) \left[1 + \frac{\dot p_{0\mu}\dot p_0^\mu}{\mathcal{F}^{2}} \left(1+2\textcolor{black}{\beta_{\mathrm{gup}}}\,p_0^\rho p_{0\rho}\right)\right].
\end{equation}
Here, $\textcolor{black}{\beta_{\mathrm{gup}}}$ is the RGUP deformation parameter. Expanding this perturbatively, the deformed metric and its inverse are given by \citep{carroll2019spacetime}:
\begin{equation}\label{r1}
    \tilde g_{\mu\nu} = \left(\delta^\alpha_{\mu} + A^\alpha_{\mu}\right) g_{\alpha\nu}, \qquad 
    \tilde g^{\mu\nu} = g^{\mu\nu} - \Xi(p_0)\frac{\dot p_{0}^{\mu}\dot p_{0}^{\nu}}{\mathcal{F}^2},
\end{equation}
where $\Xi(p_0)$ encodes the leading RGUP and curvature corrections. The determinant of the metric acquires a first-order correction \citep{thorne2000gravitation}:
\begin{equation}\label{r2}
    \sqrt{-\tilde g} \simeq \sqrt{-g}\left[ 1 + \frac12 \Xi(p_0)\frac{\dot p_{0\lambda}\dot p_{0}^{\lambda}}{\mathcal{F}^2} \right].
\end{equation}

These geometric deformations modify the canonical scalar–field action. It is important to note that the scalar field presented here is not taken to be a fundamental inflationary inflaton. Rather, as is customary in quantum-corrected cosmological models, it is used as a generic effective matter degree of freedom propagating on the RGUP-deformed spacetime background. Any effective matter Lagrangian gains corrections since RGUP alters the metric that matter fields experience. Therefore, the underlying inflationary dynamics are still controlled by the geometric $\alpha Q^2$ term in the $f(Q,L_m)$ gravity sector, and the scalar field action is only used as a convenient and model-independent probe to extract the RGUP-induced modifications to the effective energy density, pressure, and inflationary observables. By substituting the deformed metric quantities into the standard action \citep{birrell1984quantum}:
\begin{equation}\label{r3}
    S_{\varphi} = -\int d^4x\,\sqrt{-\tilde g} \left[ \frac12 \tilde g^{\mu\nu}\partial_{\mu}\varphi\partial_{\nu}\varphi + V(\varphi) \right],
\end{equation}
we obtain a corrected action split into the standard General Relativity part and the quantum correction terms:
\begin{align}
    S_{\varphi} &= -\int d^4x\,\sqrt{-g} \left[ \frac12 g^{\mu\nu}\partial_\mu\varphi\partial_\nu\varphi + V(\varphi) \right] \nonumber \\
    &\quad -\int d^4x\,\sqrt{-g} \left[ \frac12\Delta^{\mu\nu}\partial_\mu\varphi\partial_\nu\varphi +\frac12\Delta_g\left( g^{\mu\nu}\partial_\mu\varphi\partial_\nu\varphi + 2V(\varphi) \right) \right],
\end{align}
where the correction tensors are defined as:
\[
\Delta^{\mu\nu} = -\Xi(p_0)\frac{\dot p_{0}^{\mu}\dot p_{0}^{\nu}}{\mathcal{F}^2}, \qquad \Delta_g = \frac12\Xi(p_0)\frac{\dot p_{0\lambda}\dot p_{0}^{\lambda}}{\mathcal{F}^2}.
\]
Since the scalar field is introduced only as an effective description of the quantum-corrected matter sector under RGUP deformation, rather than as a fundamental inflaton that drives inflation, the potential $V(\varphi)$ is kept completely general in this case.

Consequently, the effective Lagrangian $L_m$ of the scalar field is modified by terms of order $\mathcal{O}(\mathcal{F}^{-2})$ but the geometric sector is dictated by the same $f(Q,L_m)$ action. These corrections alter the energy density $\rho_\varphi$ and pressure $p_\varphi$, which in turn modify the background dynamics via the $f(Q,L_m)$ field equations.

\subsection{Derivation of RGUP-Corrected Effective Energy Density and Pressure}

To study the cosmological implications, we must derive the effective energy density and pressure of the scalar field under the RGUP deformation. Starting from the modified action as given in Eqn. \eqref{r3}, we substitute the first-order expansions derived in the previous section:
\begin{equation}
    \sqrt{-\tilde{g}} \approx \sqrt{-g}(1 + \Delta_g), \qquad \tilde{g}^{\mu\nu} \approx g^{\mu\nu} + \Delta^{\mu\nu},
\end{equation}
where $\Delta_g = \frac{1}{2}\Xi \frac{\dot{p}^2}{\mathcal{F}^2}$ and $\Delta^{\mu\nu} = -\Xi \frac{\dot{p}^\mu \dot{p}^\nu}{\mathcal{F}^2}$.

Expanding the action to linear order in the deformation parameters, we define an effective Lagrangian density $L_{eff}$ \citep{mukhanov2005physical}:
\begin{align}\label{r4}
    S_{\varphi} &= - \int d^4x \sqrt{-g} (1 + \Delta_g) \left[ \frac{1}{2} (g^{\mu\nu} + \Delta^{\mu\nu}) \partial_\mu \varphi \partial_\nu \varphi + V(\varphi) \right] \nonumber \\
    &\approx \int d^4x \sqrt{-g} \left[ -\frac{1}{2} (g^{\mu\nu} + \Delta^{\mu\nu} + \Delta_g g^{\mu\nu}) \partial_\mu \varphi \partial_\nu \varphi - (1 + \Delta_g) V(\varphi) \right].
\end{align}
For a spatially flat FLRW background, the scalar field is homogeneous, $\varphi = \varphi(t)$, implies that $\partial_\mu \varphi \partial_\nu \varphi \rightarrow \dot{\varphi}^2 \delta_\mu^0 \delta_\nu^0$. Using the metric signature \textcolor{black}{$(-,+,+,+)$}, we have $g^{00} = -1$. It is convenient to include all the corrections to the kinetic sector brought on by the RGUP into a dimensionless quantity \textcolor{black}{$\delta_{\mathrm{RGUP}}$} that measures the deviation of the scalar sector from canonical form.
Thus, we have:
\begin{equation}\label{r5}
    \textcolor{black}{\delta_{\mathrm{RGUP}}} = \Delta_g + \Delta^{00} = \frac{1}{2}\Xi(p_0)\frac{\dot{p}^2}{\mathcal{F}^2} - \Xi(p_0)\frac{(\dot{p}^0)^2}{\mathcal{F}^2}.
\end{equation}

The effective Lagrangian for the inflaton then simplifies to:
\begin{equation}\label{r6}
    \textcolor{black}{L_{\mathrm{eff}}} = \frac{1}{2}(1 - \textcolor{black}{\delta_{\mathrm{RGUP}}}) \dot{\varphi}^2 - (1 + \Delta_g) V(\varphi).
\end{equation}
Using the standard definitions \citep{mukhanov2005physical} for the energy-momentum tensor components, $p =\textcolor{black}{L_{\mathrm{eff}}}$ and $\rho = \frac{\partial L_{\mathrm{eff}}}{\partial \dot{\varphi}}\dot{\varphi} - L_{\mathrm{eff}}$, we obtain the RGUP-corrected equations of state:
\begin{align}
    \textcolor{black}{p_{\varphi}^{\mathrm{RGUP}}}  &= \frac{1}{2}(1 - \textcolor{black}{\delta_{\mathrm{RGUP}}}) \dot{\varphi}^2 - (1 + \Delta_g) V(\varphi), \label{eq:press_rgup} \\
   \textcolor{black}{\rho_{\varphi}^{\mathrm{RGUP}}} &= \frac{1}{2}(1 - \textcolor{black}{\delta_{\mathrm{RGUP}}}) \dot{\varphi}^2 + (1 + \Delta_g) V(\varphi).
\end{align}
These modified forms of $\rho$ and $p$ include quantum-geometric corrections and act as a source in the Friedmann equations of $f(Q,Lm)$, whereas the geometric sector is unaltered.

\section{RGUP-Corrected Inflationary Results}\label{sec8}

The deformation in the metric under RGUP affects the kinetic as well as the potential terms in the inflaton, as we have seen in the previous section. This affects the inflation scenario through the slow roll parameters, depending on the normalization of the kinetic terms as well as the form of the effective potential. As a result, there are small but nonzero corrections in the scalar spectral index, tensor scalar ratio, and the running. The scalar field introduced in the RGUP framework should not be viewed as a new fundamental inflaton field, but rather as an effective description of the quantum-corrected matter sector propagating on the same $f(Q,L_m)$ geometric background. The quadratic non-metricity term $\alpha Q^2$ still governs the underlying inflationary dynamics, whereas RGUP only introduces perturbative quantum-geometric corrections to the inflationary observables, pressure, and effective energy density.

Instead of re-deriving the full slow-roll formalism, we can utilize some controlled, but phenomenological, parameterization \citep{mukhanov2005physical, liddle2000cosmological} that can capture the leading RGUP effects in the highly curved region. With this, we can write the inflationary observables in terms of some perturbative deviations around a Starobinsky form \citep{starobinsky1980new}.
Thus, to quantify the impact of the quantum geometric corrections derived in Section \ref{sec7}, we parameterize the deformation terms using a dimensionless parameter $\textcolor{black}{\beta_{\mathrm{gup}}}$. The effective modification to the spectral index and tensor-to-scalar ratio is modeled as $n_s \approx n_s^{(0)} + \textcolor{black}{\beta_{\mathrm{gup}}}/N$ and $r \approx r^{(0)}(1-2\textcolor{black}{\beta_{\mathrm{gup}}})$. At this point in our study, we note that this parametrization can be understood as a first-order phenomenological expansion around the background, which is Starobinsky-like, in which the metric deformation induced by the RGUP rescales the kinetic part of the effective scalar field. This approach provides a concise description of the leading-order impact of Planck-scale corrections in scenarios in which the full perturbation equations are analytically unsolvable.
In this case, the values of $n_s^{(0)}$ and $r^{(0)}$ correspond to the predictions of the inflationary scenario, which can be obtained from the high-curvature limit of the $f(Q, L_m)$ model, whereas the RGUP parameter $\textcolor{black}{\beta_{\mathrm{gup}}}$ accounts for quantum-geometry corrections.

\subsection{Running of the Spectral Index}
Higher-order quantities offer a more sensitive probe of subleading dynamical effects \citep{aghanim2020planck} since the inflationary observables $n_s$ and $r$ are already tightly constrained and show strong degeneracies among a broad class of inflationary models. Specifically, it is known that changes in the kinetic structure and higher-derivative corrections in the inflationary scenario particularly affect the running of the scalar spectral index and the running of the running.
They are, therefore good diagnostic tools for isolating RGUP-induced effects.

In our study, the RGUP deformation modifies the effective slow-roll parameters through a modified normalization of the inflaton kinetic term. As a result, tiny but finite corrections are added to the scalar perturbations' scale dependence. The running of the spectral index, denoted by \( \alpha_s \) and the second running denoted by \( \beta_s \), are taken to be derivatives with respect to comoving wave number $k$ in the standard inflationary scenario \citep{liddle2000cosmological, aghanim2020planck, mukhanov2005physical}
\begin{align}
\alpha_s &\equiv \frac{d n_s}{d\ln k} \simeq -\frac{d n_s}{dN} = \frac{\beta_{\text{gup}} - 2}{N^2},\\[4pt]
\beta_s &\equiv \frac{d \alpha_s}{d\ln k} \simeq -\frac{d \alpha_s}{dN} = \frac{2\,(\beta_{\text{gup}} - 2)}{N^3}.
\end{align}
where N is the number of e-folds before the end of inflation.
We perform a numerical analysis for $N=60$ e-folds and compare the standard $f(Q,L_m)$ model ($\textcolor{black}{\beta_{\mathrm{gup}}}=0$) against RGUP-deformed scenarios with $\textcolor{black}{\beta_{\mathrm{gup}}} = 0.05$.  We also include a number of representative inflationary models that are frequently discussed in the literature for a wider context, such as Hilltop/D-Brane models \citep{kallosh2019hilltop}, natural inflation \citep{freese1990natural}, and chaotic inflation \citep{linde1982new}. The results are summarized in Table \ref{tab2}.

\begin{table}[!htbp]
\centering
\caption{Comparison of inflationary observables and running parameters at $N=60$ e-folds. We compare the RGUP model against the standard Starobinsky attractor \cite{starobinsky1980new}, Chaotic inflation \citep{linde1982new}, Natural inflation \citep{freese1990natural}, and Hilltop/D-Brane models \citep{kallosh2013superconformal}. The RGUP model predicts a distinct running value while maintaining a viable tensor-to-scalar ratio.}
\label{tab2}

\resizebox{\textwidth}{!}{
\begin{tabular}{l c c c c}
\toprule
\textbf{Model} & $\mathbf{n_s}$ & $\mathbf{r}$ & $\boldsymbol{\alpha_s}$ & $\boldsymbol{\beta_s}$ \\
\midrule
\textbf{Chaotic} ($\phi^2$) & 0.9667 & 0.1333 & $-0.00056$ & $-0.00002$ \\
\textbf{Natural} ($\cos(\phi/f)$) & $\approx 0.9600$ & $\approx 0.0500$ & $-0.00062$ & $-0.00003$ \\
\textbf{Hilltop / D-Brane} ($p=4$) & 0.9500 & $< 10^{-4}$ & $-0.00083$ & $-0.00003$ \\
\textbf{Starobinsky} ($\textcolor{black}{\beta_{\mathrm{gup}}}=0$) & 0.9667 & 0.0033 & $-0.00056$ & $-0.00002$ \\
\midrule
\textbf{RGUP} ($\textcolor{black}{\beta_{\mathrm{gup}}}=0.05$) & \textbf{0.9675} & \textbf{0.0030} & $\mathbf{-0.00054}$ & $\mathbf{-0.000018}$ \\
\bottomrule
\end{tabular}
}
\end{table}

A critical question in quantum-deformed cosmologies is whether the introduced corrections represent real physical effects or just a reparameterization of the classical inflationary background. A sensitive theoretical diagnostic of such effects can be obtained in the current framework by running the spectral index. We note that the RGUP deformation causes a non-vanishing shift in the running parameter, $|\Delta \alpha_s| = |\alpha_s^{\mathrm{RGUP}} - \alpha_s^{\mathrm{Standard}}| \approx 2 \times 10^{-5}$ which indicates an imprint of higher-order quantum-geometric corrections to the inflationary observables.
This deviation shows that the momentum-dependent metric deformation introduces a principled departure from the standard Starobinsky attractor at the theoretical level, even though it is far below the current sensitivity of Planck 2018 ($\sigma(\alpha_s) \approx 0.007$) \citep{meerburg2020planck}. Since the corrections are small enough to maintain the successful scale-invariant predictions of the $f(Q,L_m)$-driven inflationary background while still being distinct enough to form a non-trivial quantum-geometric signature, the RGUP-corrected scenario thus occupies a phenomenologically favorable regime. This behavior demonstrates that the RGUP framework is not merely a mathematical artefact but rather a consistent and physically significant extension of the classical inflationary scenario.

\begin{table}[htbp]
\centering
\caption{Comparison between the $f(Q,L_m)$ inflationary model and the RGUP-corrected scenario in the high-curvature regime.}
\label{tab3}
\begin{tabularx}{\textwidth}{l X X}
\toprule
\textbf{Feature} & \textbf{$f(Q,L_m)$ (High Curvature)} & \textbf{$f(Q,L_m)$ + RGUP Corrections} \\
\midrule
Gravitational action & $f(Q) \simeq -Q + \alpha Q^2$ & Same geometric action (unchanged) \\
Inflation driver & Quadratic non-metricity term $\alpha Q^2$ & Same (geometric origin preserved) \\
Role of matter sector & Negligible ($L_m \approx 0$) & Modified effective $L_m$ via RGUP deformation \\
Background dynamics & Starobinsky-like attractor & Same background evolution (perturbatively corrected) \\
Scalar spectral index $n_s$ & $n_s^{(0)} = 1 - \frac{2}{N}$ & $n_s \approx n_s^{(0)} + \frac{\beta_{\text{gup}}}{N}$ \\
Tensor-to-scalar ratio $r$ & $r^{(0)} = \frac{12}{N^2}$ & $r \approx r^{(0)}(1 - 2\beta_{\text{gup}})$ \\
Running $\alpha_s$ & Standard slow-roll value & Slight RGUP-induced shift \\
Limit $\beta_{\text{gup}} \to 0$ & --- & Reduces exactly to baseline $f(Q,L_m)$ inflation \\
Physical interpretation & Purely geometric inflation & Quantum-geometric perturbative extension \\
\bottomrule
\end{tabularx}
\end{table}

Table \ref{tab3} highlights the structural differences between the $f(Q,L_m)$ inflationary scenario and the RGUP corrected model in the high-curvature limit. From the table, we can see that the geometric part of the action and the $\alpha Q^2$ driving inflation remain unchanged, whereas the deformation of the theory under the influence of the RGUP mainly affects the matter part of the action. The cosmological evolution thus retains its Starobinsky-like character, whereas the inflationary predictions attain a small perturbative correction, ensuring consistency with the original geometric inflationary scenario.

\subsection{Comparison with Planck 2018 Constraints}
\begin{figure}[htbp]
    \centering
    \includegraphics[width=0.9\linewidth]{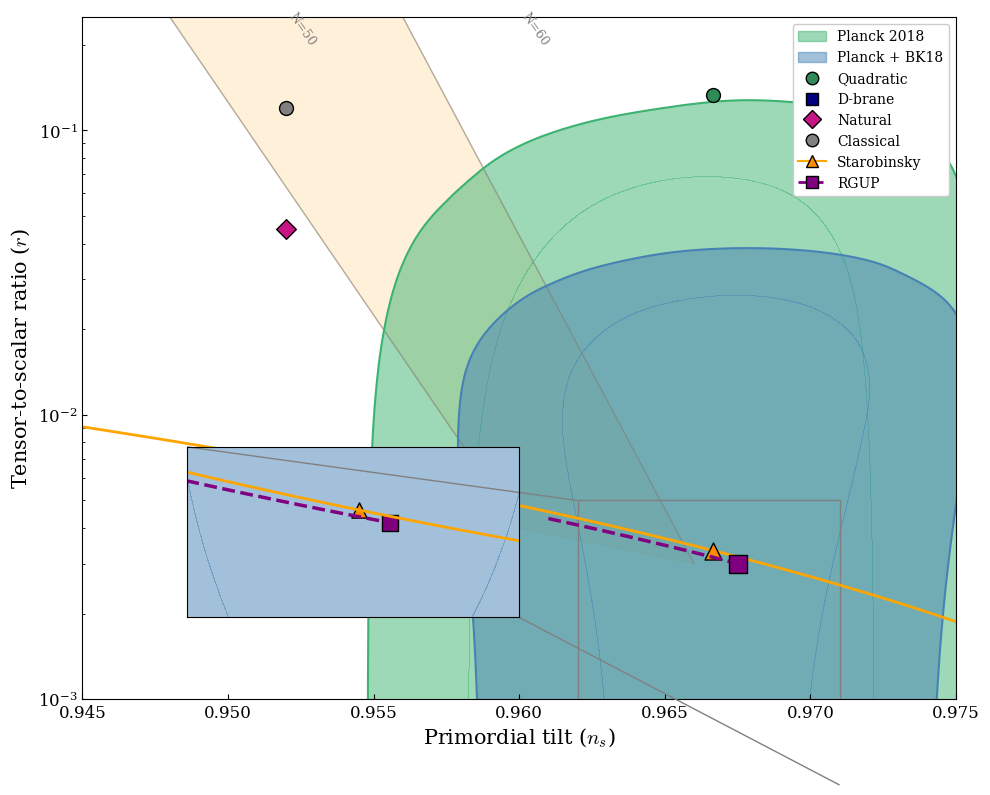}
  \caption{Constraints on $n_s$ and $r$ from Planck 2018 and BK18 (68\% and 95\% CL). We compare the RGUP model (purple square, $\beta=0.05$) against standard Starobinsky inflation (orange triangle) and other benchmark models at $N=60$. The inset highlights the suppression of the tensor ratio in the RGUP model relative to the standard Starobinsky attractor.}
    \label{f5}
\end{figure}

The theoretical predictions are visualized in Fig. \ref{f5}. We note that the Planck 2018 \citep{aghanim2020planck} and BK18 \citep{ade2022latest} confidence contours in the figure are generated using publicly available likelihood contour data sets based on MCMC chains obtained from the LAMBDA data archive. These data sets are used for comparison with the theoretical predictions of the model for the observables consistency test. The RGUP correction shifts the observables slightly towards a higher spectral index and lower tensor-to-scalar ratio, maintaining excellent agreement with the Planck 2018 confidence regions.

The RGUP correction causes a slight shift toward a slightly larger $n_s$ and suppressed tensor-to-scalar ratio in comparison to the standard Starobinsky attractor, as seen in Fig.\ref{f5}, while staying well inside the permitted Planck confidence regions. This deviation in the region where the Starobinsky-like models cluster is highlighted in the inset. It shows that the RGUP deformation introduces a theoretically distinct prediction in $n_s-r$ plane that is fully consistent with current data, even though the effect is currently undetectable. This supports the idea that RGUP is not a phenomenological contradiction but rather a controlled quantum-geometric extension of the $f(Q,L_m)$ inflationary background.

\section{Conclusion}\label{sec9}
In this study, we examined a non-minimal matter–geometry coupling in $f(Q,L_m)$ gravity in order to study a unified cosmological scenario within the framework of symmetric teleparallel gravity. We showed that the same underlying geometric theory can explain both early-time inflation and late-time cosmic acceleration through a natural separation of curvature scales by using a minimal but physically well-motivated polynomial form of the gravitational Lagrangian, 
\begin{equation}
  f(Q,L_m) = -Q + \alpha Q^2 + 2L_m + \beta QL_m \nonumber
\end{equation}
The background expansion at late times is controlled by the matter–geometry coupling term proportional to $\beta$. Holographic Ricci Dark Energy was used as an effective fluid to reconstruct the cosmological dynamics, which leads to the derivation of a nonlinear evolution equation (Eqn. \eqref{17}) for the evolution of the Hubble parameter, including the effects of non-metricity corrections and matter-geometry coupling. The graphs obtained from the numerical reconstruction are shown in Fig.\ref{f1}. We observe that the evolution history is smooth and stable, with the Hubble parameter monotonically decreasing, the deceleration parameter remaining negative, and the scale factor undergoing constant acceleration. Additionally, the EoS is dynamically evolving, showing the transition from the quintessence regime to the phantom regime, finally approaching the de Sitter regime (see Fig.\ref{f2}), which is evidence for the validity of the model in describing the evolution of the universe at late times under the joint effect of non-metricity and holographic corrections.

As inferred from Fig.\ref{f3}, current late-time observations only place a one-sided constraint on the coupling parameter $\beta$, with no statistical preference for a non-zero value, according to a Bayesian analysis using Pantheon supernovae, cosmic chronometers, and DESI 2024 BAO data. This behavior, which is entirely consistent with the $\Lambda$CDM limit, shows an extended degeneracy between the background expansion history and mild matter–geometry couplings. Crucially, as anticipated, given its function in the early Universe, the high-energy parameter $\alpha$ is unconstrained by late-time data.
The quadratic non-metricity term $\alpha Q^2$ dominates the gravitational action while the matter contributions and the matter-geometry coupling become insignificant in the high- curvature context that is relevant to inflation. The model naturally reduces to a Starobinsky-like inflationary scenario in this limit, driven solely by geometric effects, producing predictions for the spectral index ($n_S$) and tensor-to-scalar ratio ($r$) that are consistent with Planck 2018 constraints (see Fig.\ref{fig:inflation_ns_r}).
Thus, we observe that rather than necessitating distinct mechanisms, inflation and late-time acceleration potentially emerge as distinct limits of the same gravitational theory.
We investigated the robustness of the inflationary predictions under quantum-gravity-inspired corrections, driven by the extreme energy scales typical of inflation. We demonstrated that quantum-geometric effects modify the effective inflaton energy density and pressure without changing the classical $f(Q,L_m)$ background dynamics by implementing RGUP through a momentum-dependent deformation of the spacetime metric. Higher-order inflationary observables, especially the running of the scalar spectral index, are slightly but finitely altered by these corrections as seen in Table \ref{tab2}. The resulting deviations show that RGUP effects cause a small but non-negligible departure from the classical Starobinsky attractor while maintaining observational viability, even though they are below the sensitivity of current CMB observations. This has also been pictorially depicted in Fig. \ref{f5}.

At this point, we would like to briefly emphasize how the present study differs from the foundational work done by Hazarika \textit{et.al.} in \citep{hazarika2024f}, where the authors primarily focused on establishing the theory and its cosmological implications in general. On the other hand, our study incorporates a specific minimal polynomial model along with HRDE and observational datasets to explore the phenomenology across different curvature regimes. In addition, the RGUP-inspired quantum corrections also extend the classical framework. Thus, this work can be seen as an extension and complement of the original theory as it provides a common geometric framework wherein the inflation and late-time acceleration arise from different curvature regimes.

Possible extensions of the current analysis could be to consider more general forms of $f(Q,L_m)$, extending the minimal quadratic polynomial form adopted in this work to other, more general forms that could include higher-order and non-polynomial non-metricity terms, as well as more general forms of matter-geometry couplings. This could potentially lead to an evaluation of the model dependence of the inflation-dark energy unification scenario in the context of systematic teleparallel gravity. Another possible direction is to extend the analysis to the perturbation level, which could include the evolution of scalar and tensor perturbations and the CMB power spectra, potentially leading to a more robust investigation of the validity of the RGUP correction scheme.

\section*{Declaration of generative AI and AI-assisted technologies in the writing process}

During the preparation of this work the author(s) used Grammarly and Quillbot in order to improve the language and correct the grammar. After using this tool/service, the author(s) reviewed and edited the content as needed and takes full responsibility for the content of the publication.

\section*{Data Availability Statement}
The Type Ia supernova data used in this work come from the Pantheon data compilation, which is publicly available at \url{https://github.com/dscolnic/Pantheon}.\\ 
The BAO constraints were taken from the results for the DESI DR1 BAO cosmology, which is publicly available at \url{ https://data.desi.lbl.gov/public/dr1/vac/dr1/bao-cosmo-params}
the Planck 2018 $r$–$n_s$ confidence contours were obtained from\\
\url{https://github.com/kdemirel/Planck-constraints-r-vs-ns-plot}.

\bibliographystyle{unsrtnat}
\bibliography{r.bib}
\end{document}